\documentclass{emulateapj}
\pdfoutput=1
\usepackage{natbib}
\usepackage{times}
\usepackage{graphicx}
\usepackage{amsmath}
\newcommand\jcap{{J. Cosmology Astropart. Phys.}}%
\newcommand\pasa{{PASA}}%
  
\begin{document}

{\vspace*{-0.8cm}\hfill LA-UR-12-10321,~UCB-NPAT-12-004,~NT-LBNL-12-003}

\shorttitle{Cen~A UHECR Excess and Local Ext. Magnetic Field}
\shortauthors{ Y{\"U}KSEL, STANEV, KISTLER, \& KRONBERG}
\title{The Centaurus A Ultrahigh-Energy Cosmic Ray Excess\\ and the Local Extragalactic Magnetic Field}
\author{Hasan Y{\"u}ksel\altaffilmark{1},
Todor Stanev\altaffilmark{2},
Matthew D. Kistler\altaffilmark{3,4,5},
and Philipp P. Kronberg\altaffilmark{1,6}}

\altaffiltext{1}{Theoretical Division, MS B285, Los Alamos National Laboratory, Los Alamos, NM 87545}
\altaffiltext{2}{Bartol Research Institute, Department of Physics and Astronomy, University of Delaware, Newark, DE 19716}
\altaffiltext{3}{Lawrence Berkeley National Laboratory and Department of Physics, University of California, Berkeley, CA 94720}
\altaffiltext{4}{California Institute of Technology, MC 350-17, Pasadena, CA 91125}
\altaffiltext{5}{Einstein Fellow}
\altaffiltext{6}{Department of Physics, University of Toronto, Toronto M5S 1A7, Canada}


\begin{abstract}
The ultrahigh-energy cosmic-ray anisotropies discovered by the Pierre Auger Observatory give the potential to finally address both the particles' origins and properties of the nearby extragalactic magnetic field (EGMF).  We examine the implications of the excess of $\sim 10^{20}\,$eV events around the nearby radio galaxy Centaurus~A.  We find that, if Cen~A is the source of these cosmic rays, the angular distribution of events constrains the EGMF strength within several Mpc of the Milky Way to $\gtrsim\,20\,$nG for an assumed primary proton composition.  Our conclusions suggest that either the observed excess is a statistical anomaly or the local EGMF is stronger than conventionally thought.  We discuss several implications, including UHECR scattering from more distant sources, time delays from transient sources, and the possibility of using magnetic lensing signatures to attain tighter constraints.
\end{abstract}
\preprint{LA-UR-12-10321, UCB-NPAT-12-004, NT-LBNL-12-003}


\section{Introduction}
Significant progress has been made toward determining the origin of ultrahigh-energy cosmic rays (UHECRs; see \citealt{Hillas:1985is} and \citealt{Beatty:2009zz,Kotera:2011cp,LetessierSelvon:2011dy} for recent reviews), with the Pierre Auger Observatory (Auger) acquiring data needed to achieve this and other goals.  However, a definitive identification of sources remains elusive, as early indications of a correlation of UHECR events with AGN within $\sim$100~Mpc from the Veron-Cetty and Veron catalog~\citep{Cronin:2007zz} have given way to a less-clear, yet still anisotropic, picture~\citep{Abreu:2010zzj}.

\begin{figure}[b!]
\hspace*{-0.5cm}
\includegraphics[width=1.1\columnwidth,clip=true]{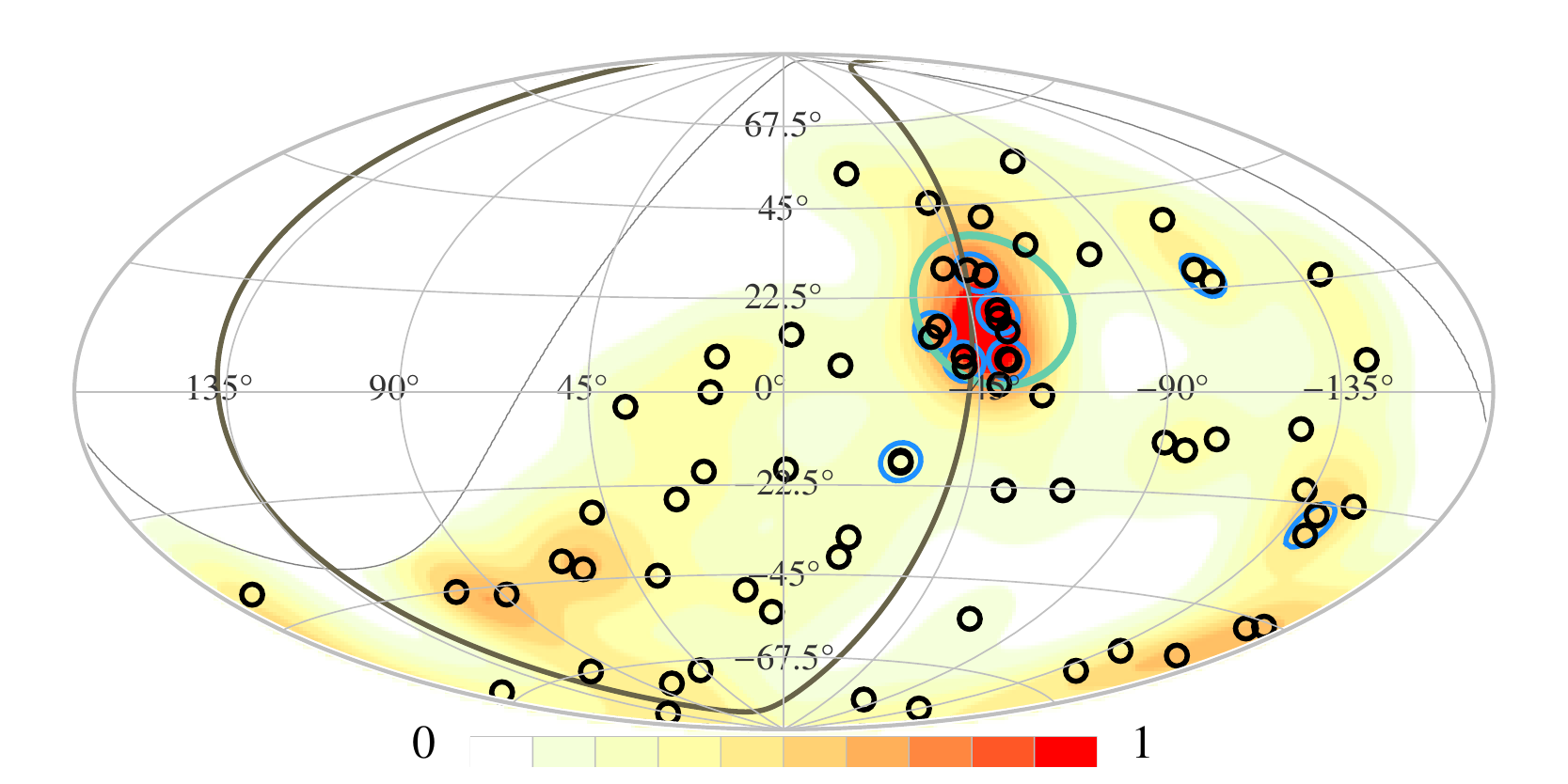}
\caption{The arrival directions of 69 UHECR events detected by Auger with $E \geq 5.5 \times 10^{19}\,$eV ({\it black circles}) in Galactic coordinates~\citep{Abreu:2010zzj}.  Event pairs within $5^\circ$ of each other are further circled.  The light solid line demarcates the horizon of Auger, while the darker solid line corresponds to the supergalactic plane.  We also show an estimated density distribution ({\it contours}) obtained by replacing discrete events with a Gaussian and weighting by the relative exposure of Auger.  A circle of $18^\circ$ radius is drawn around the center of the radio galaxy Cen~A.
\label{augersky}}
\end{figure}

The most prominent feature in the UHECR sky is a significant excess of events from the vicinity of Centaurus~A (Cen~A; see Fig.~\ref{augersky}) \citep{Abraham:2007si,Abreu:2010zzj,Gorbunov:2007ja,Stanev:2008sd,Moskalenko:2008iz,Hillas:2009yh}, a nearby active radio galaxy possessing well-studied giant radio lobes~\citep{Junkes1993,Feain:2009rf,Feain:2011as}.  These radio lobes, along with the proximity to the Milky Way, have long made Cen~A a prime prospective source of UHECR~\citep{Cavallo1978,Romero:1995tn,Ahn:1999jd,Farrar:2000nw,Isola:2001ng}.  This is in stark contrast to the lack of an excess towards the Virgo galaxy cluster, home of the powerful AGN M87.

A major difficulty in tracing cosmic rays, even those with energies exceeding $10^{20}\,$eV, back to their sources is the uncertain nature of both the Galactic magnetic field (GMF)~\citep{Stanev:1996qj} and the extragalactic magnetic field (EGMF) in the vicinity of the Milky Way~\citep{Kronberg:1993vk,Widrow:2002ud}.  The EGMF remains poorly constrained, especially locally.  While upper limits of $\sim\,$1-10~nG with coherence lengths of $\sim\,$1-50~Mpc have been placed in cosmological settings~\citep{Blasi:1999hu}, a global estimate of the energy output of all supermassive black holes implies $\sim\,$0.1-1~$\mu$G fields spread out within the filaments connecting clusters of galaxies~\citep{Colgate:2011enout}.  Faraday rotation studies suggest fields as large as  $\sim\,$0.3~$\mu$G in the filaments~\citep{Xu:2005rb} and  $\sim\,$1-10~$\mu$G in intra-cluster environments~\citep{Clarke:2000bz}.  A hope of UHECR studies is to increase our knowledge of the properties of the EGMF (e.g., ~\citealp{Kronberg:1993vk,Lee:1995tm,Sigl:1998dd,Stanev:2000fb,Dolag:2004kp,Anchordoqui:2001bs,Anchordoqui:2001nt,Armengaud:2004yt,Ryu:2009pf,Jiang:2010yc,Anchordoqui:2011ks,Takami:2012uw}).  One way to do so requires evidence that a known object is indeed producing UHECRs in order to reconstruct the intervening field structure.

We examine the implications of the Cen~A excess by comparing the observed distribution of $> 5.5 \times 10^{19}\,$eV Auger events with the expectations from UHECR propagation in a turbulent EGMF.  Using this technique, we find that the overall angular distribution can be well reproduced for a range of magnetic field configurations if Cen~A is an UHECR source.  This then allows us to constrain the strength and the structure of the local EGMF.  Our focus is on the effects of the nearby EGMF on the propagation of UHECRs, and later we address why the GMF should not impact our conclusions.

Throughout this study, we assume that Cen~A is the source of the excess, and that the UHECR are protons.  We will discuss alternatives, while addressing Auger results~\citep{Abreu:2011vm} that do not display the excess around Cen~A that is expected at lower energies, if the high-energy events are heavy nuclei~\citep{Lemoine:2009pw}.  The observed rapid isotropization with decreasing energy is most easily interpreted as proton domination and arises in subtle ways in a number of our viable field configurations.  We develop further immediate implications of our scenario, including the effect on angular distributions from more distant sources, a requisite minimum time delay for UHECRs from transient sources (e.g., gamma-ray bursts), and the possibility of using magnetic lensing effects to narrow the allowed magnetic field parameter space.

\begin{figure}[b!]
\includegraphics[width=\columnwidth,clip=true]{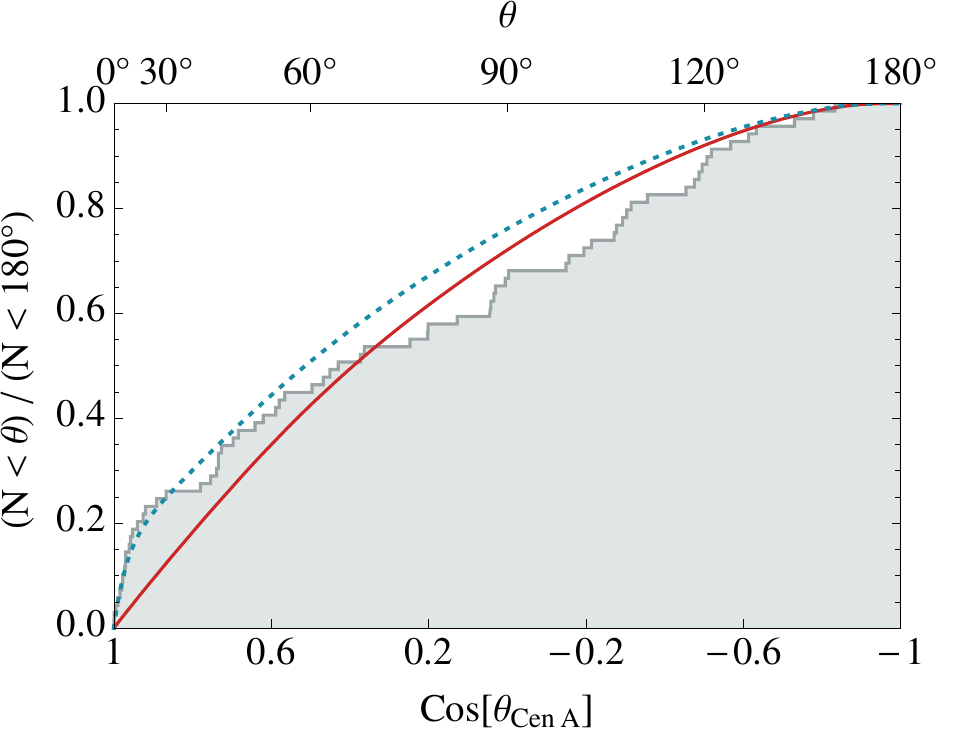}
\caption{The cumulative angular distribution of the 69 Auger UHECR events when centered on Cen~A ({\it steps}).  
We compare, after weighting for exposure, the expectations for: ({\it i}) a purely isotropic distribution of all events ({\it solid}), and; ({\it ii}) 10 events following a $10^\circ$ Gaussian distribution around Cen~A upon an otherwise isotropically distributed 59 events ({\it dotted}). 
\label{cenacum}}
\end{figure}

\section{Cen~A's Place in the UHECR Sky}
The deflection of UHECRs during propagation, which depends on their charge/momentum and any magnetic fields encountered, has long frustrated attempts at associating observed events with well-known astronomical objects.  Owing to their extreme energies, $\gtrsim 10^{20}$~eV cosmic-ray protons should not experience strong deflections in typical Galactic ($\sim \mu$G) magnetic fields, and may point back to near their birthplaces if extragalactic fields  are not too large~\citep{Stanev:2000fb}.  A corollary to this is that the UHECR distribution from a known source can be used to determine the structure and the strength of the magnetic fields.

We show in Fig.~\ref{augersky} the 69 UHECR events with $E \geq 5.5 \times 10^{19}\,$eV ({\it dots}) reported through March 2009~\citep{Abreu:2010zzj}.  In order to better visually represent the UHECR density distribution, we have also replaced each event with a Gaussian~\citep{Silverman} of width $10^\circ$ (comparable to the spread of excess events around Cen~A).  To correct for the non-uniform exposure of Auger, we also inversely re-weight each event using the exposure function~\citep{Sommers:2000us} at its location in the sky, so that events with lower exposure display higher significance.   The resulting density distribution is normalized to unity and is displayed as shaded contours.

It is clearly apparent that no other strong excess exists other than in the direction of Cen~A.  Quantitatively, 13 events are observed within 18$^\circ$ of Cen A, while $\sim$3 events are expected on average for a purely isotropic distribution~\citep{Abreu:2010zzj}, suggesting an excess of about $\sim$10 events.  In Fig.~\ref{cenacum}, we show the cumulative distribution of the Auger events in terms of the angular separation from Cen~A (using Cos$\,\theta_{\rm Cen A}$ for uniform coverage in solid angle), along with the expectations, after weighting with the Auger exposure, for: ($i$) an isotropically distributed set of events (solid line); ($ii$) a $10^\circ$ Gaussian containing 10 events centered at Cen~A superimposed on an otherwise isotropic distribution (dotted line).

The \citet{Abreu:2010zzj} determined that only 4\% of random realizations of 69 events drawn from an isotropic distribution deviate from the assumption of isotropy by more than the data itself do.  Following this prescription, we find $\sim$6\% deviations for model ($i$), in good agreement with the \citet{Abreu:2010zzj}, and $\sim$8\% for model ($ii$).  This suggests that even accounting for an excess from the direction of Cen~A does not alter the fact that the all-sky distribution of events is anisotropic.  Indeed, examining Fig.~\ref{augersky} we see a large low-density void, where the exposure is nearly maximal, that rivals the Cen~A excess in prominence.

The projected angular distribution is part of a larger story.  Closer examination reveals that many of these events are contained in clusters, and further that most of the clusters are near Cen~A.  In Fig.~\ref{augersky}, we put small circles around the events within $5^\circ$ of each other. Additionally, the events are not symmetric about the position of Cen~A.  These peculiarities may result from the particular structure of the EGMF, a possibility that we later address.

There are a number of possibilities for accelerating UHECR in a radio galaxy like Cen~A \citep{Kronberg:2004acc,Hardcastle:2008jw,Fraschetti:2008uc,OSullivan:2009sc,Dermer:2008cy,Rieger:2009pm,Pe'er:2011xf,Sahu:2012wv}.  Acceleration in the inner jets or near the black hole both would occur on a scale much smaller than the radio lobes.  It is also possible that particle acceleration occurs within the giant lobes themselves.  For our purposes, we assume that for all instances particle injection arises from a central point source, since the apparent extent of the excess exceeds even the $\sim 8^\circ \times 2^\circ$ angular size of the lobes.

\section{Magnetic Field Formulation}
Any treatment of cosmic ray propagation in the nearby universe requires a description of the magnetic field structure of the local EGMF.  To simulate this, we generate a divergence-free, random magnetic field model whose components have a Gaussian distribution and follow $\left| B_{k}\right|^2 \propto k^{-(n+2)}$. Here $n=5/3$ corresponds to Kolmogorov turbulence.  Adopting the prescription detailed in~\citet{Tribble:1991xx} and~\citet{Murgia:2004zn}, we first construct Fourier components of a complex valued vector potential ${\mathbf A}$ whose components follow a power spectrum of the form $\left| A_{k}\right|^2 \propto k^{-(n+4)}$ in a 3D cubical box in wave number (${\mathbf k}$) space.  For a given ${\mathbf k}$, each component $A_i$ is drawn from the distribution
\begin{equation}
  P(A_i,\phi)\,dA_i\,d\phi = \frac{A_i}{\left| A_{k}\right|^2}\exp\left(-\frac{A_i^2}{2 \left| A_{k}\right|^2} \right) \, dA_i \, \frac{d\phi}{2\pi}.
\label{eq:power}
\end{equation}
Solving for $A_i$, when the power spectrum normalization is omitted, yields
\begin{equation}
  A_i = \sqrt{ k^{-(n+4)}  \ln[1/\rho_1^2]} \,e^{i\, 2 \pi \, \rho_2}\,,
\label{eq:component}
\end{equation}
in which $\rho_1$ and $\rho_2$ are real random numbers in the range 0 to 1.  Construction of each ${\mathbf A}$ vector thus requires six random numbers. The values of ${\mathbf A}$ are calculated for all ${\mathbf k}$ in the range $k_{\rm min}= 2\pi\,/\ell_{\rm max} \leq  | k |  \leq 2\pi\,/\ell_{\rm min}=k_{\rm max}$, in which $\ell_{\rm min}$ and $\ell_{\rm max}$ correspond to the minimum and maximum scales of the magnetic field.  The magnetic field in ${\mathbf k}$-space is given as
\begin{equation}
  {\mathbf B}(\mathbf k) = i \, {\mathbf k} \times {\mathbf  A}(\mathbf k)\,,
\label{eq:component2}
\end{equation}
and is transformed back into coordinate space through a 3D complex fast Fourier transform.

\begin{figure}[t!]
\includegraphics[width=\columnwidth,clip=true]{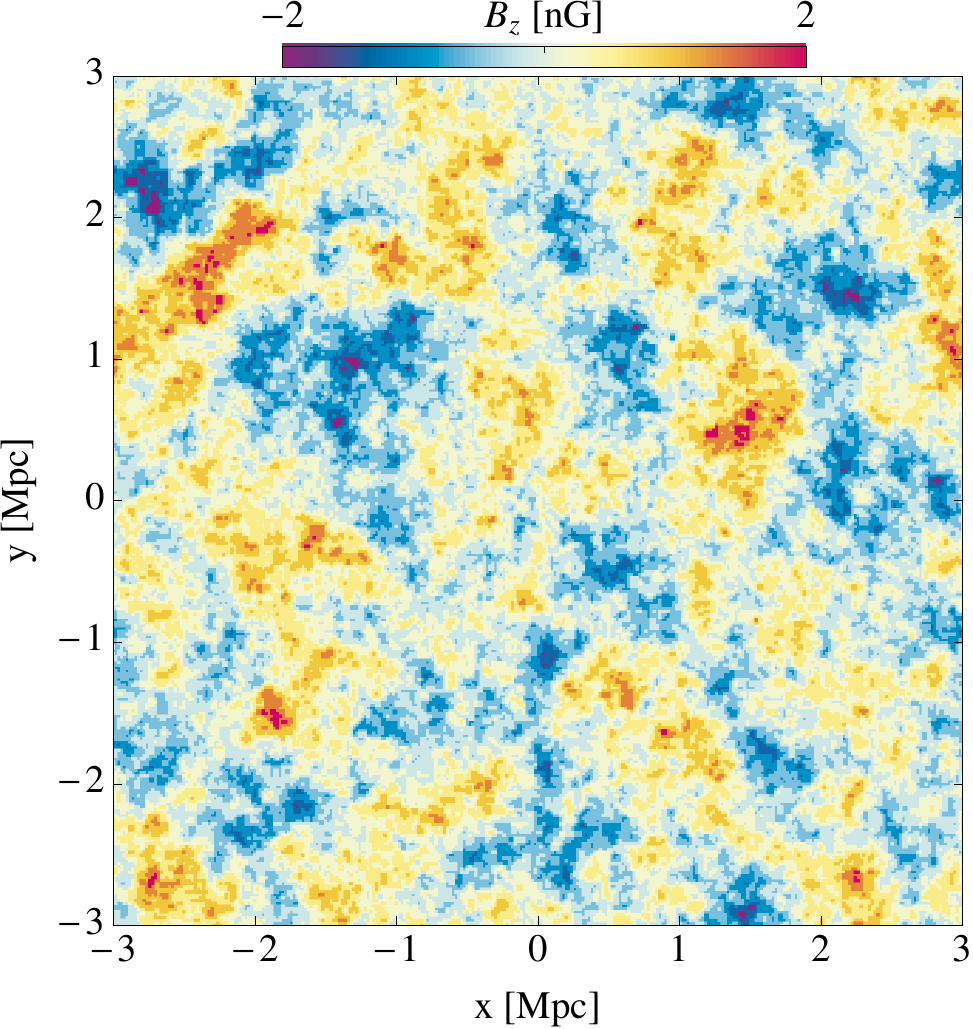}
\caption{A slice from the magnetic field configuration, obtained from a Kolmogorov spectrum within a $512^3$ cubic grid, used in this study.  Shown is the $z$ component of the field projected onto the $x$-$y$ plane.\\
\label{fig:magran}}
\end{figure}

Fig.~\ref{fig:magran} shows one slice from the magnetic field configuration obtained using a cubic grid of size $512^3$. The $z$ component of the field is projected in the $x$-$y$ plane.  Here, we choose $k_{\rm min}= 3 $, $k_{\rm max} =256$ for a Kolmogorov spectrum (although ${\mathbf A}$ for $k_{\rm max} > 128$ is assumed to be 0 as the size of the grid will be too coarse to properly resolve them and these very small scale fluctuations would be under sampled).  The quadratic mean of the magnetic field is normalized to $B_{\rm rms}=1$~nG over the volume and the wave numbers are scaled such that $\ell_{\rm max}=2$~Mpc (for $k=3$) and $\ell_{\rm min} \simeq 0.04$~Mpc (for $k=128$) in coordinate space.

The most relevant quantities to describe any configuration of the magnetic field structure are $B_{\rm rms}$ and the coherence length $\Lambda_c$, which is defined~\citep{Harari:2002dy} as
\begin{equation}
  \Lambda_c \simeq \frac{\ell_{\rm max}}{2}\frac{n-1}{n}\frac{1-(\ell_{\rm min}/\ell_{\rm max})^n}{1-(\ell_{\rm min}/\ell_{\rm max})^{n-1}}\,.
\end{equation}

When we use this magnetic field configuration to track the trajectories of charged particles for a given field strength ($B_{\rm rms}$) and coherence length ($\Lambda_c$), we renormalize $B_{\rm rms}$ and rescale $\ell_{\rm min}$ and $\ell_{\rm max}$ in our magnetic field simulation box accordingly.  Choosing $k_{\rm min}= 3$ ensures that there is sufficient variation in the simulation box even at the largest scales.  We take this field to be static over the relatively short propagation times of the particles.  The simulation box is periodically repeated when the propagation distance exceeds the box size.

\section{Obtaining Angular Distributions}
In order to understand the angular distribution of events seen by Auger, we first look for a range of EGMF parameters that can result in a spread of $\sim 10^\circ$ for UHECR arriving from Cen~A.  We will use analytical expressions for guidance and later compare to in-depth simulations.  In our equations, unless stated otherwise, we express both distance and time in Mpc (since $v\simeq c$ for ultra-relativistic particles), energy in EeV ($10^{18}$ eV), magnetic field strength in nG, and consider singly-charged particles, applicable for protons.  The equation of motion is governed by the Lorentz force, which, in a region lacking electric fields, can be cast as a set of first order coupled equations
\begin{equation}
  \frac{d \, {\boldsymbol \beta}}{d\,t} \simeq 0.925 \, \frac{{\boldsymbol  \beta}\times {\bf B}}{E} \hspace*{1cm} {\boldsymbol  \beta} = \frac{d \,{\bf r}}{d\,t}\,,
\label{eq:eom}
\end{equation}
where we express the velocity vector in terms of $c$, thus, ${\boldsymbol \beta}$ is a unit vector following these equations.  The Larmor radius in a constant magnetic field is given as
\begin{equation}
  r_L \simeq 1.08\, {E}/{B} \,.
  \label{eq:larrad}
  \end{equation}

While the energy of the particle in these equations can be replaced with the rigidity $R=E\,/\,Z$ for a nucleus with charge $Z$, we keep this implicit and simply use $E$ rather than $R$ unless otherwise noted.  All of our subsequent results are quoted for a given energy and can be scaled from a proton to an iron primary by accounting for the appropriate charge through either the normalization of magnetic field or energy, although care must be taken in reinterpreting our ultimate conclusions for the case of heavy nuclei, as we discuss later.  Due to the proximity of Cen~A, we do not include energy losses here.

In a turbulent magnetic field with a given strength, $B_{\rm rms}$, and coherence length, $\Lambda_c$,  the quadratic mean of the scattering of the final particle velocity with respect to the initial velocity is
\begin{equation}
  \delta_{\rm rms} \simeq\,53^{\circ} \,\sqrt{1/2} \, B_{\rm rms}\, \sqrt{d \, \Lambda_c}\,/ E \,,
\label{eq:scat}
\end{equation}
where $d$ is the distance travelled \citep{Harari:2002dy}.  The corresponding scatter in arrival directions of particles around the source as seen at Earth is
\begin{equation}
  \theta_{\rm rms} = \delta_{\rm rms} /\sqrt{3} \,.
\label{eq:scat2}
\end{equation}

While Eq.~(\ref{eq:scat}) is valid for $d \gg \Lambda_c$, when $d \ll \Lambda_c$ the particle effectively travels in a domain in which the magnetic field is constant. In that case, the deflection could be larger and will appear as a shift from the position of the source rather than scattered around the source.  In this case, the change in the direction of the particle velocity, when averaged over all magnetic field configurations, can be expressed as
\begin{equation}
  \delta_{\rm av} \simeq\, 53^{\circ}  \,\sqrt{2/3} \,  B_{\rm rms} \,d\, / E \,,
\label{eq:bend}
\end{equation}
in which the factor $\sqrt{2/3}$ accounts for the average perpendicular component of the random magnetic field direction with respect to the velocity. In this case, the shift in the observed arrival directions around the source is
\begin{equation}
  \theta_{\rm av} = \delta_{\rm av} /2 \,.
\label{eq:bend2}
\end{equation}
One can combine these two relations as
\begin{eqnarray}
  \theta& \simeq &\left(\theta_{\rm av}^{\,\eta}+\theta_{\rm rms}^{\,\eta}\right)^{1/\eta} \nonumber\\
    &\simeq& \,\,53^{\circ}\,\sqrt{1/6}\, B_{\rm rms}\, (d/{E}) \left(({\Lambda_c}/{d})^{\eta/2}+1 \right)^{1/\eta}  ,
\label{eq:comboscat}
\end{eqnarray}
which describes either the average shift or the quadratic mean angular scattering of particles from the source. Here $\eta \rightarrow -4$ parametrizes the smoothness of the transition between the two regions in which $\Lambda_c \gg d$ and $\Lambda_c \ll d$.

\begin{figure}[t]
\includegraphics[width=0.49\textwidth,clip=true]{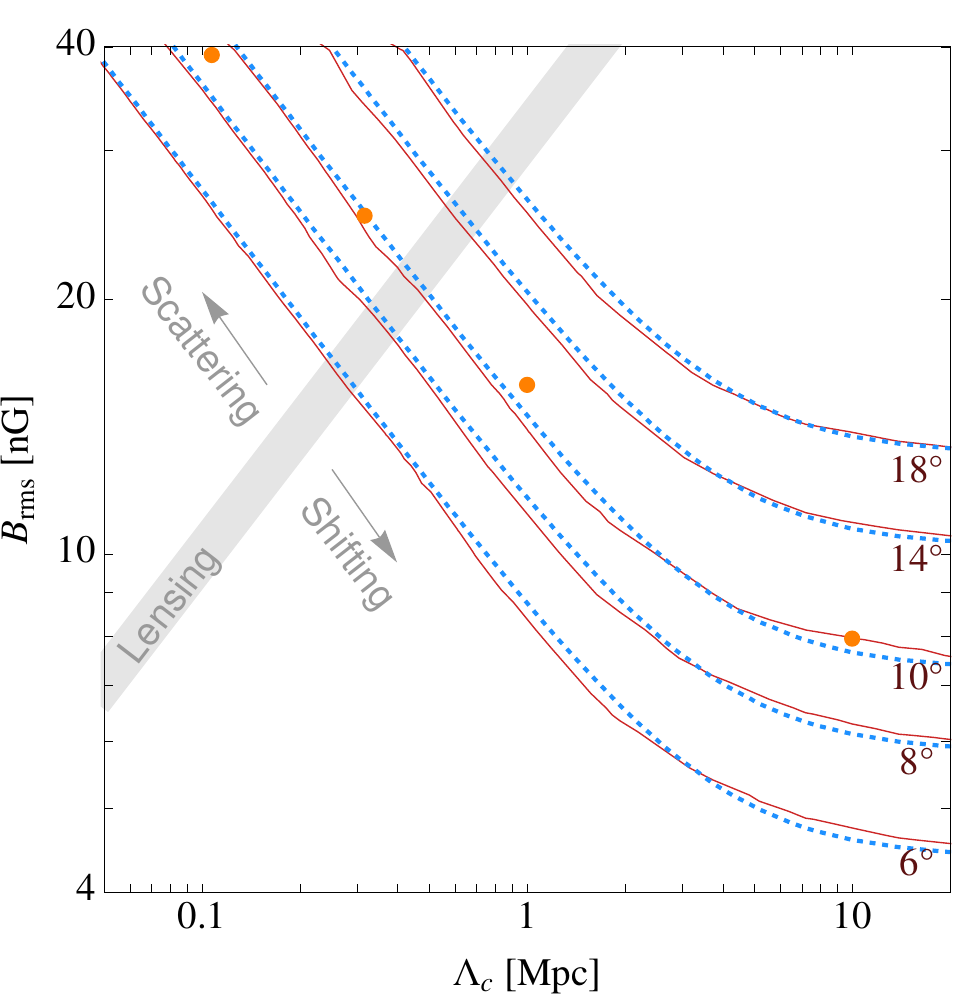}
\caption{The average angular distribution of 60~EeV cosmic-ray protons around Cen~A.  Shown are the expectations from Eq.~(\ref{eq:comboscat}) ({\it dotted lines}) and the results of simulations through turbulent magnetic fields ({\it solid lines}).  We identify regions where the behavior is dominated by either scattering ($\Lambda_c \ll d$) or shifting ($\Lambda_c \gg d$) of the source flux as seen by an observer.  Magnetic lensing effects are strongest along the gray band (see text).  Detailed results of simulations are presented for parameters denoted by large orange dots in Fig.~\ref{fig:sky}.\\
}
\label{fig:delta-sim}
\end{figure}

We consider particles with $E=60$~EeV, comparable to the majority of the Auger events, propagating from a distance to Cen~A of 3.8~Mpc~\citep{Harris:2009wj}.  Fig.~\ref{fig:delta-sim} shows values for $\theta$ (dotted lines) based on Eq.~(\ref{eq:comboscat}) for a given $\Lambda_c$ and $B_{\rm rms}$. We next compare this analytic result with numerical simulations.

We compute $\theta$ utilizing a fourth-order Runga-Kutta method to solve Eq.~(\ref{eq:eom}), keeping the step size small in comparison to both the minimum scale of variation in magnetic field and $r_L$.  In order to ensure that  our results are always an average of many distinct realizations of the magnetic field configuration, we have varied the relative locations of the source and detector while keeping their geometrical arrangements fixed when necessary.

As seen in Fig.~\ref{fig:delta-sim}, the agreement between numerical results (solid lines) with the analytical results (dotted lines) is excellent.  However, these results only provide the average of the distribution, losing important details from the azimuthal distribution for comparison with the Auger data.  Fig.~\ref{fig:delta-sim} will be used as a guide as we pursue this direction via simulation.

\section{Simulations \& Scenarios}
One can identify various realms, two of which have already been mentioned and labelled in Fig.~\ref{fig:delta-sim}.  {\it Scattering} is characterized by Eqs.~(\ref{eq:scat}) \& (\ref{eq:scat2}) and corresponds to a random walk around the source position for small angles.  Larger scattering optical depth eventually leads to diffusive behavior.  {\it Shifting} describes the case when the coherence length of the field exceeds the distance to the source, so that the particles see an approximately uniform field.  This results in an apparent change in the position of the source as described by Eqs.~(\ref{eq:bend}) \& (\ref{eq:bend2}).

{\it Lensing} refers to the magnification and/or appearance of multiple images of the cosmic-ray source due to uncorrelated deflections in the magnetic field. It roughly corresponds to the transition between the two realms mentioned above.  This is an attractive possibility for using the observed angular distribution to infer greater details about the intervening magnetic field.  For distant sources, the lensing effects are strongest near the critical energy,
\begin{equation}
E_c \simeq\, 0.6\, B_{\rm rms}\, {d^{3/2}}/{\sqrt{\Lambda_c}} \,,
\label{eq:lens}
\end{equation}
as given in \citet{Harari:2002dy}.  The gray band in Fig.~\ref{fig:delta-sim} corresponds to the values of $B_{\rm rms}$ versus $\Lambda_c$ that produce the most prominent lensing from a source at a distance of $d \sim 3.8$~Mpc for particles of $E \simeq E_c/\sqrt{6} \simeq 60$~EeV.  Below this band, shifting dominates and we only have a single image, while above scattering dominates.  Around this band, the source could be magnified, which may contribute to the excess of events from the direction of Cen~A.

\begin{figure}[b!]
\includegraphics[width=\columnwidth,clip=true]{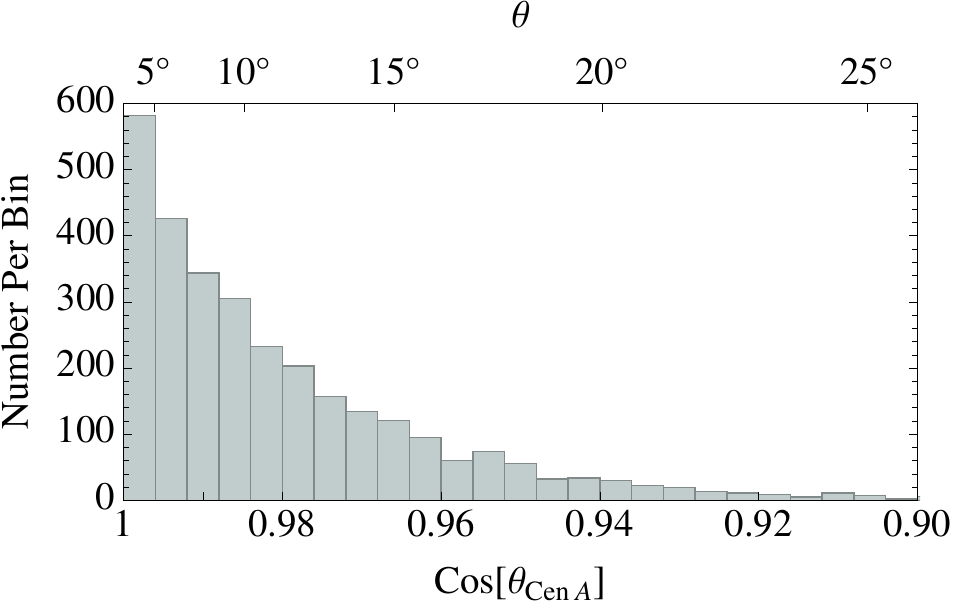}\\
\includegraphics[width=\columnwidth,clip=true]{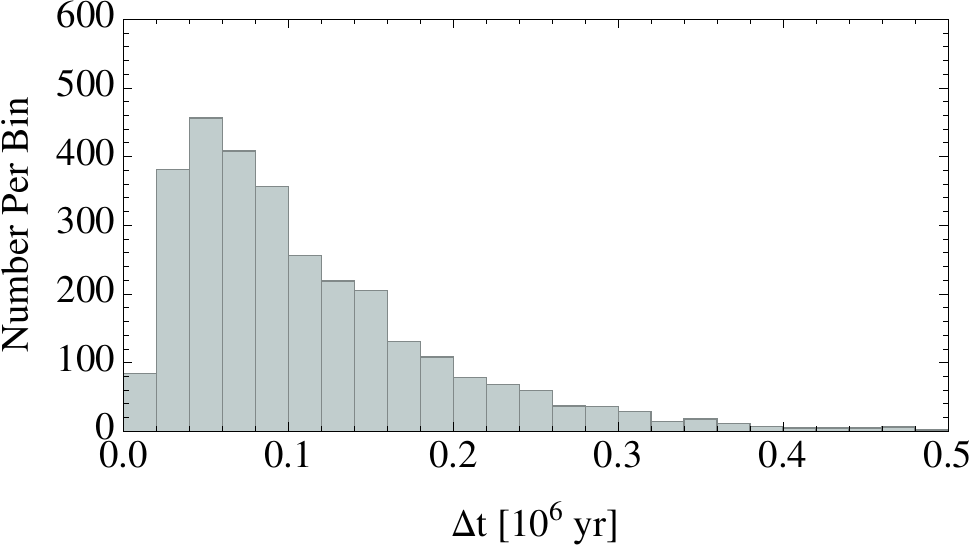}
\caption{
{\it Top:}  The observed angular distribution of 60~EeV cosmic-ray protons for EGMF parameters $B_{\rm rms} = 16$~nG and $\Lambda_c=1$~Mpc, which yields an average displacement of $\sim 10^\circ$ from Cen~A.  {\it Bottom:} The corresponding distribution of delay times (propagation time minus $d/c$), averaging $\sim\,$0.1 Myr.  All EGMF parameters that are compatible with the UHECR distribution around Cen~A give similar results.
\label{delays}}
\end{figure}

\begin{figure*}[t!]
\includegraphics[width=\columnwidth,clip=true]{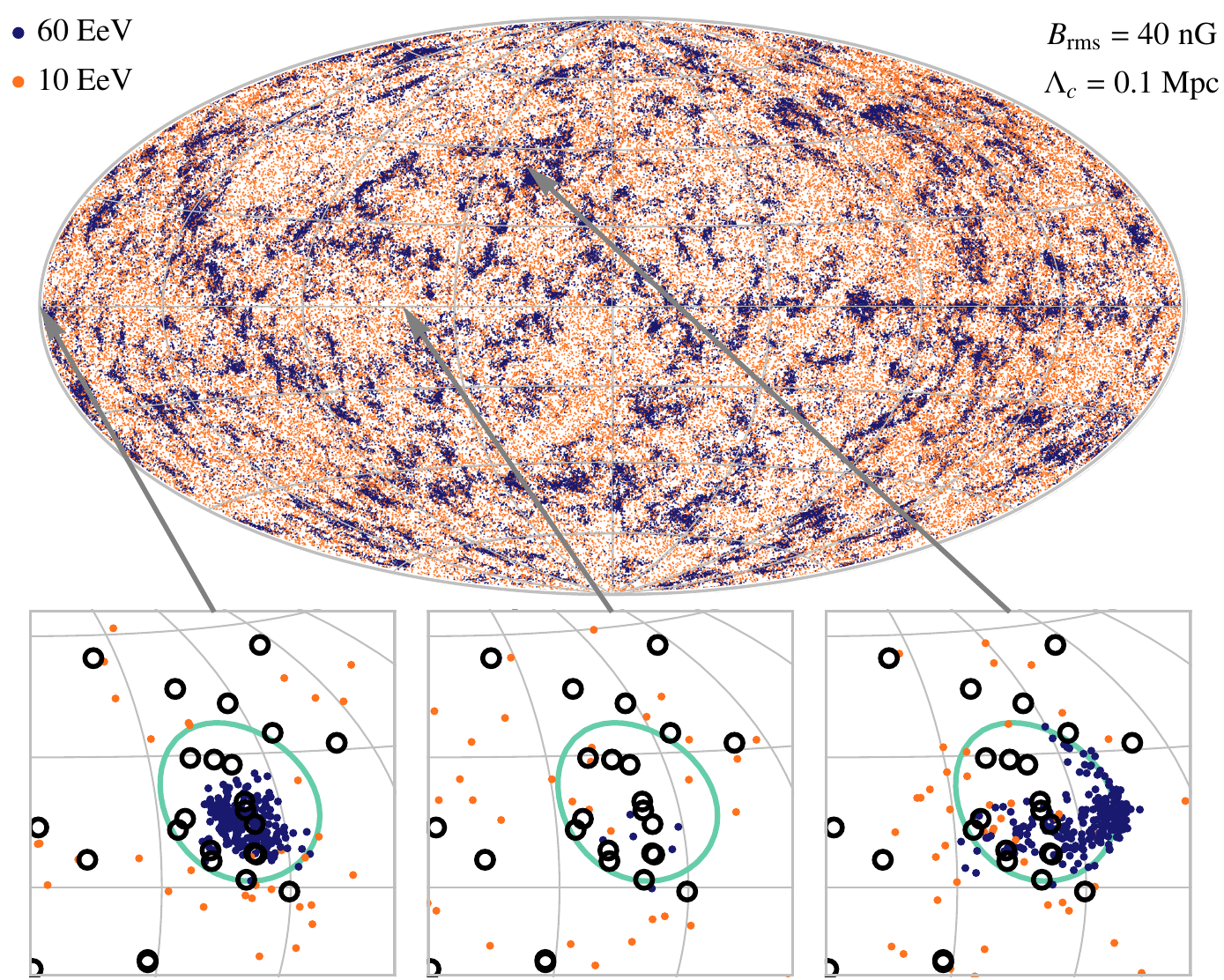}
\includegraphics[width=\columnwidth,clip=true]{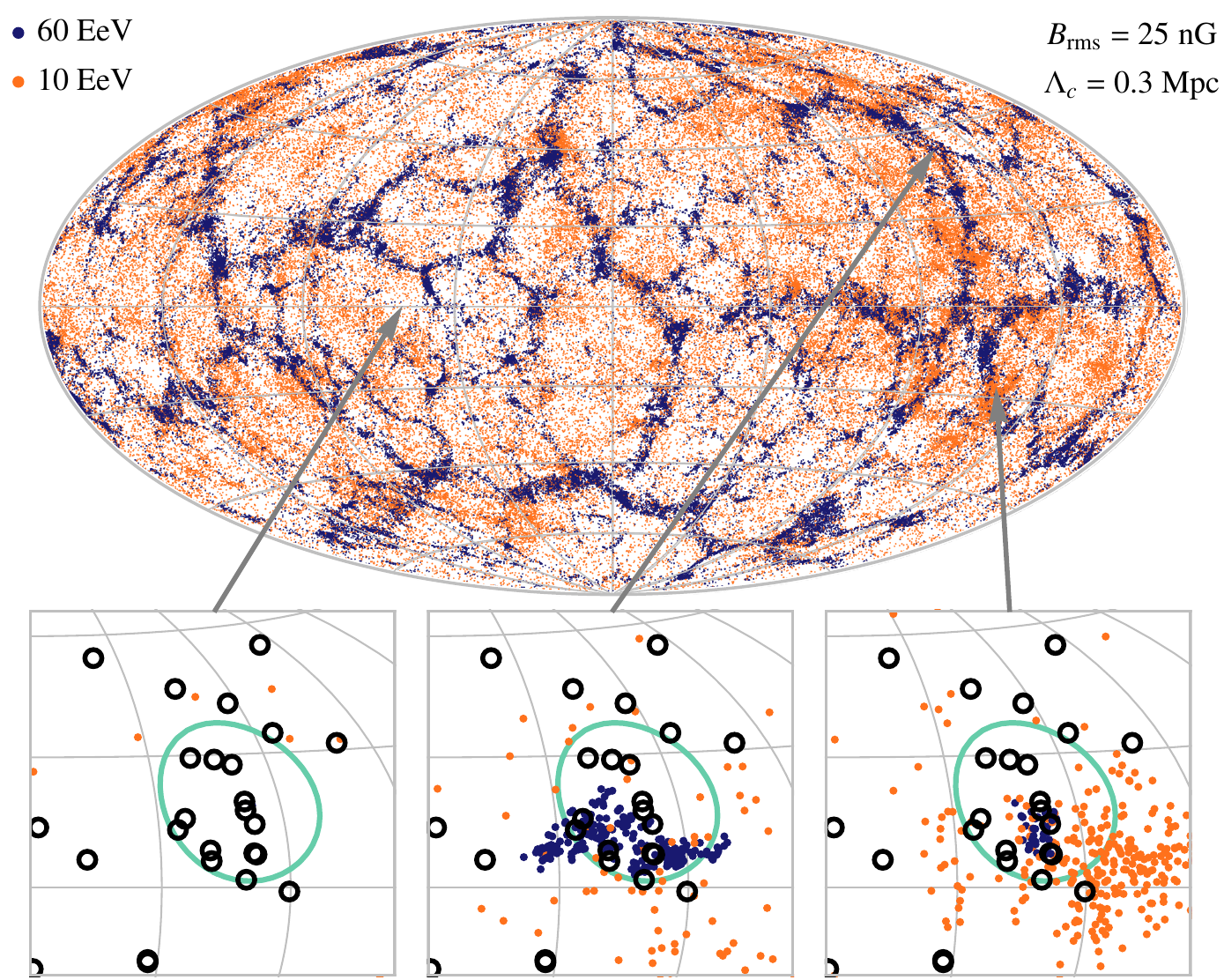}\\
\vspace*{0.5cm}
\includegraphics[width=\columnwidth,clip=true]{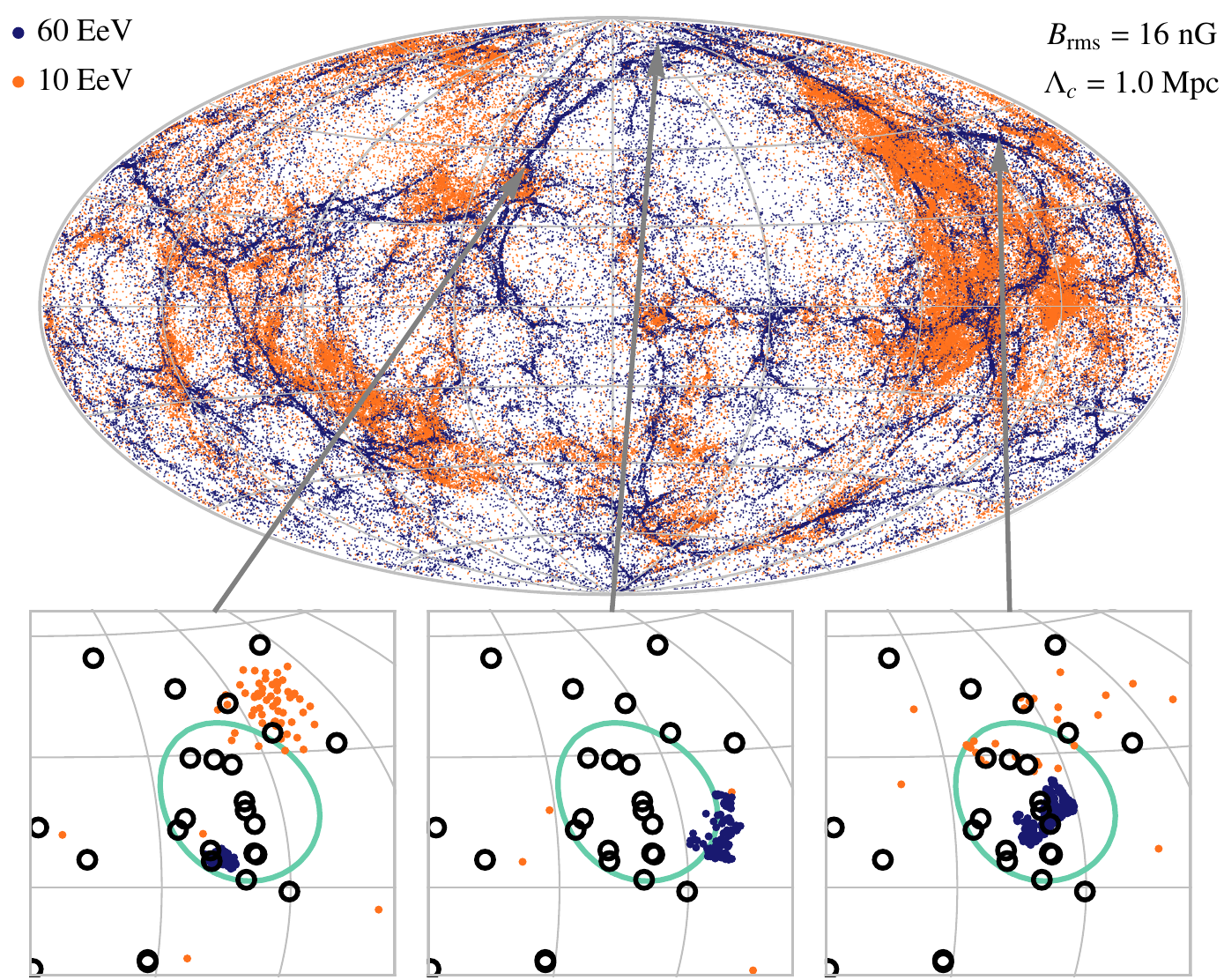}
\includegraphics[width=\columnwidth,clip=true]{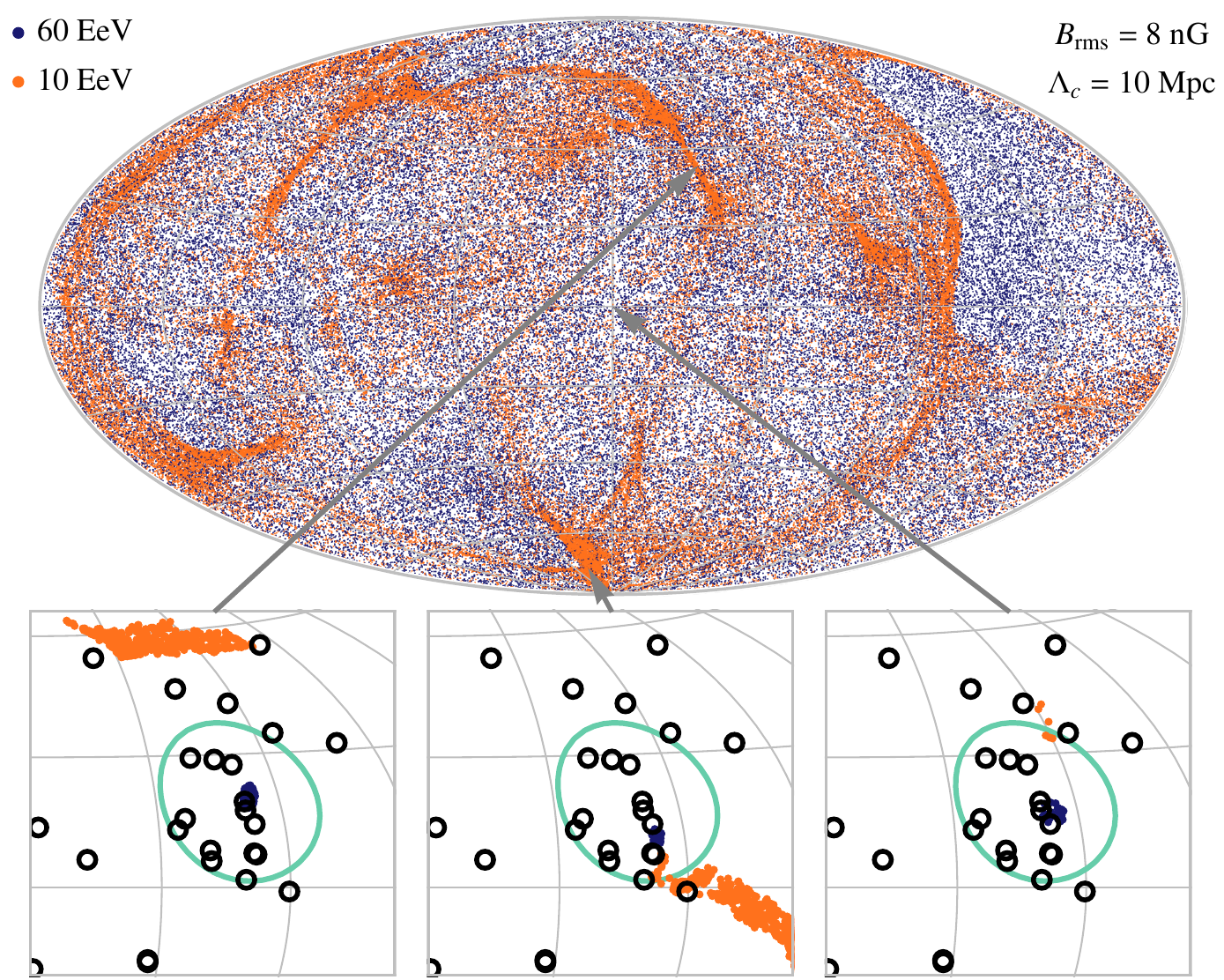}
\caption{Distributions of cosmic rays for four different extragalactic magnetic field parametrizations ($B_{\rm rms}$ vs. $\Lambda_c$) using the magnetic field structure from Fig.~\ref{fig:magran}.  Populations of protons with energies of 60~EeV ({\it dark blue points}), similar to the observed Auger event energies, and 10~EeV ({\it light orange points}) are shown.  For each of the four models:
{\it Top:} The final positions of particles, as seen by an observer at Cen~A, after reaching a distance of 3.8~Mpc.  $10^5$ particles are shown for each energy.
{\it Bottom:} Three characteristic realizations of UHECR angular distributions arriving from Cen~A, as seen at Earth, chosen from locations in the map above (as marked with arrows) shown together with Auger data ({\it black circles}).\\
\label{fig:sky}}
\end{figure*}

To better demonstrate these realms, we have chosen magnetic field parameters for UHECR propagation simulations in order to display realizations in which the all-sky averaged displacement of all trajectories from the center of Cen~A is $\sim 10^\circ$ (denoted by the large orange dots in Fig.~\ref{fig:delta-sim}).  The top panel of Fig.~\ref{delays} shows this for one such set.  In these simulations, we have used a small test sphere at a distance of 3.8~Mpc from the source on which incoming particles are detected and we make the resulting sky plots.  The average scatter ($\sim 2^\circ$) among the dark blue points in the lower right realization of Fig.~\ref{fig:sky} illustrates the effective resolution determined by the size of the sphere, chosen as a compromise between higher statistics and image quality.

We display the results of these simulations in Fig.~\ref{fig:sky}.  In the top panel of each of the four realizations, we show the all-sky distribution of UHECR (the positions of particles after reaching a sphere of radius 3.8~Mpc) as seen by an observer located at Cen~A for cosmic rays of energies 60~EeV (dark blue points), to compare with the Auger data, and 10~EeV (light orange points) in order to illustrate behavior at lower energies. In the bottom panels, we show the sky distributions seen at Earth from three different locations relative to Cen~A (as marked in the upper panels by arrows) for each simulation, along with the Auger data (black circles) for comparison. The location of Cen~A is at the center of the large green circle. These illustrate the unique features that can arise in these configurations. Despite this, the average angular spreads are roughly equal.

Starting from small coherence lengths, we see in the upper panels that the distribution of trajectories as seen from the source is rather isotropic for the 60~EeV cosmic rays, due to the scattering being the largest.  As $\Lambda_c$ is increased, large-scale clustering becomes more prominent as the lensing regime is traversed.  This gives way to a more isotropic dispersal again at large $\Lambda_c$ due to the lack of scattering.  For the 10~EeV cosmic rays, a similar track is evident, although, since the EGMF parameters were chosen based on the 60~EeV results, this is not as well defined in the figures.  For both energies, when lensing features are most prominent, significant fractions of the sky receive relatively low fluxes, so that a simple inverse-square based inversion may not yield the true luminosity of the source.

As seen from Earth, in many cases when $\Lambda_c$ is low a spread is present in the 60~EeV arrival directions, with sub-clustering among these sometimes present, along with an overall displacement from the center of Cen~A.  Conversely, larger coherence lengths yield tight clustering with an overall shift in position relative to Cen~A.  These display how the average angle of all events in the simulation can be $\sim 10^\circ$, as also given by the previous formulae, do not yet yield a distribution similar to the Auger data around Cen~A.  This permits general conclusions to be drawn regarding the properties of the EGMF despite the inherent difficulties involved with examining an infinite parameter space.

\begin{figure}[t!]
\includegraphics[width=\columnwidth,clip=true]{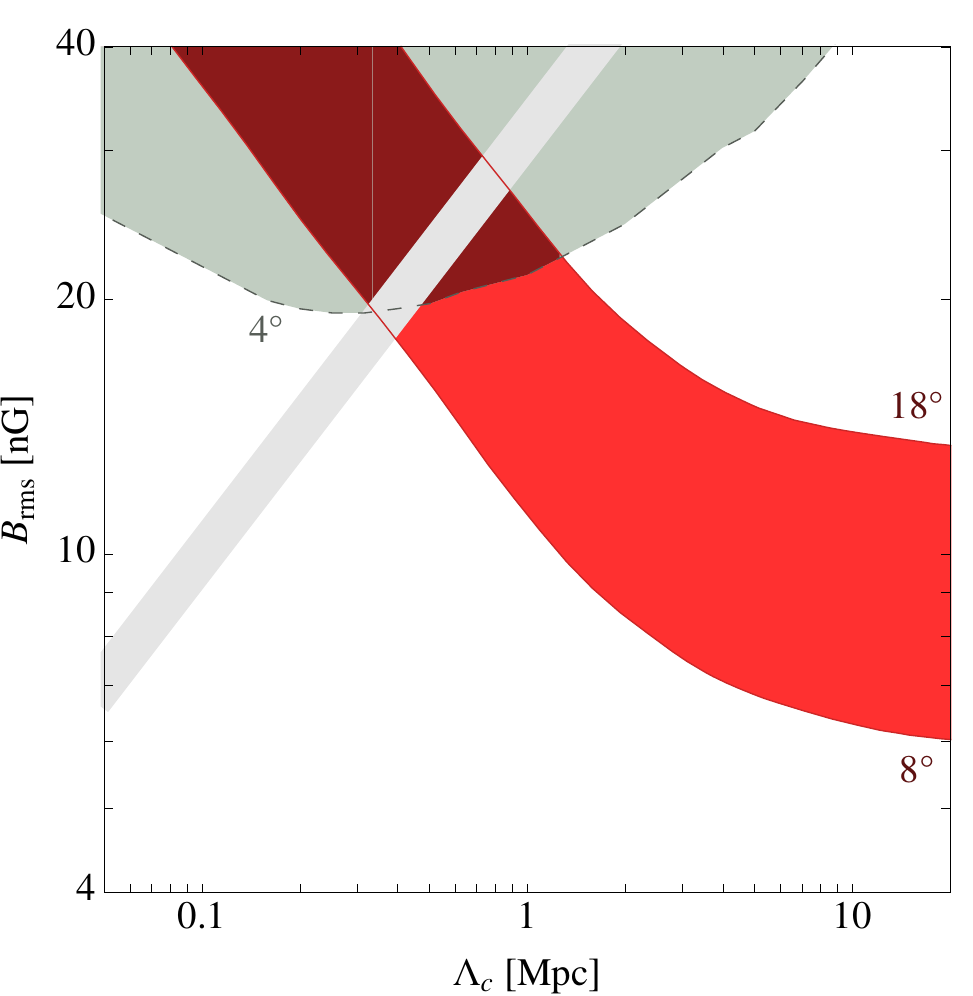}
\caption{Inferred range of extragalactic magnetic field parameters $B_{\rm rms}$ vs. $\Lambda_c$ ({\it dark shaded region}) that is compatible with both: ($i$) the average angular distribution of events being 8$^\circ$-18$^\circ$ from Cen~A (between the {\it solid lines}); ($ii$) the spread of events among themselves having an extent larger than 4$^\circ$(above the {\it dashed line}).  The latter condition disfavors scenarios in which events are shifted from the source position yet remain tightly clustered.  These conservative requirements indicate $B_{\rm rms}$ should exceed 20~nG in the local extragalactic environment if Cen~A is the source of the Auger excess.  As in Fig.~\ref{fig:delta-sim}, the gray band corresponds to the appearance of lensing effects, which would allow for tighter constraints.\\
\label{limits}}
\end{figure}

\section{The Local Extragalactic Magnetic Field}
We have assumed in the above that the EGMF is the dominant source of the scattering of UHECRs from a nearby object and now provide constraints based on observations. We address the deflections by GMF at the close of this Section.  We show in Fig.~\ref{limits} the regions in the parameter space of $B_{\rm rms}$ and $\Lambda_c$ that satisfy two conditions based on the Auger data.  First, we require the average angular distribution of events, averaged over the whole sky in order to use the highest statistics, to be 8--18$^\circ$ from Cen~A.  This yields the shaded region between the solid lines.

As discussed above, large $\Lambda_c$ values can meet this standard yet still result in point-like distributions.  We thus further require the ``internal'' spread among the arrival directions to be larger than 4$^\circ$, with our simulations resulting in the dashed lines.  The intersection of these regions results in the preferred combination of EGMF parameters (dark shaded region).  While for random field configurations, even with the same choice of parameters, it is difficult to define sharp boundaries on the allowed properties of the EGMF, we have attempted to estimate these under the assumption that Cen~A is the cause of the UHECR excess coincident with its position in the sky.  Techniques have been developed to compare in detail two-dimensional distributions of data (e.g., \citealt{Peacock1983}), although, given the present data, our aim here is to provide a broad perspective.

The most direct implication of this result is that, if Cen~A is the source of protons resulting in the excess seen by Auger, then the strength of the intervening EGMF is $\gtrsim 20\,$nG.  Measurements of the depth of maximum for UHECR showers at $\lesssim 40$~EeV made by Auger over large regions of the sky (i.e., not specifically around Cen~A), suggest a heavy nuclei composition \citep{Abraham:2010yv}.  Similar measurements from HiRes (located in the northern hemisphere and not covering Cen~A) instead indicate a light composition \citep{Abbasi:2009nf}, which leads to an ambiguous situation concerning the all-sky composition.

However, if the high-energy excess surrounding Cen~A is due to heavy nuclei, this would imply a flux of protons of the same rigidity that propagate along the same trajectories if the accelerated particles were drawn from a solar-like composition \citep{Lemoine:2009pw}.  This leads to an expected excess at lower energies of the same angular extent, though not as prominent due to the larger expected isotropic background.  This was not seen by Auger; however, leading to limits on the elemental composition to ratios that would be far from solar values if the $> 55$~EeV excess is due to heavy nuclei \citep{Abreu:2011vm}.  The simplest interpretation of this result is a dominant proton component.

This result has many consequences.  A number of studies have suggested that the filling factor of extragalactic space containing fields of at least this strength is not very large (e.g., \citealp{Sigl:2004yk,Dolag:2004kp,Das:2008vb,Takami:2008ri,Giacinti:2009fy}).  However, the Milky Way being a fairly large galaxy may give credence to our being located within a filament containing a relatively-strong field, as seen elsewhere in the nearby universe \citep{Kronberg:2007wa}.  Large scale simulations of extragalactic magnetic fields have difficulty in dealing with details at the scales of relevance here, which nonetheless can have large effects on UHECR propagation.

From the available data alone, it is not possible to infer field properties beyond our neighborhood.  However, a $\gtrsim 20\,$nG field extending at least $\gtrsim\,$Mpc around the Milky Way results in a ``screen'' scattering all UHECRs that eventually reach Earth.  Each UHECR would then be expected to have a minimum amount of deflection due to this field alone.  If this were comparable to the angular extent of the Cen~A excess, it may increase the difficultly of making associations with more distant sources.  Further, this local screen would introduce a minimum time dispersion.  To illustrate this effect, we show in Fig.~\ref{delays} our simulated increase in arrival times from Cen~A due to the EGMF, which averages to $\sim\,$0.1~Myr.  This is important for transient sources, such as gamma-ray bursts, which would experience at least this amount.

We also show, in Fig.~\ref{limits}, the band corresponding to where particles with energies $\lesssim\, 60$~EeV produce multiple images of the source.  We see that this runs orthogonal to our preferred region, so that, if the presence of multiple images can be inferred for a given range of energies using the improved statistics of future data sets, a narrow range of field properties would be established.

\begin{figure}[b]
\includegraphics[width=\columnwidth,clip=true]{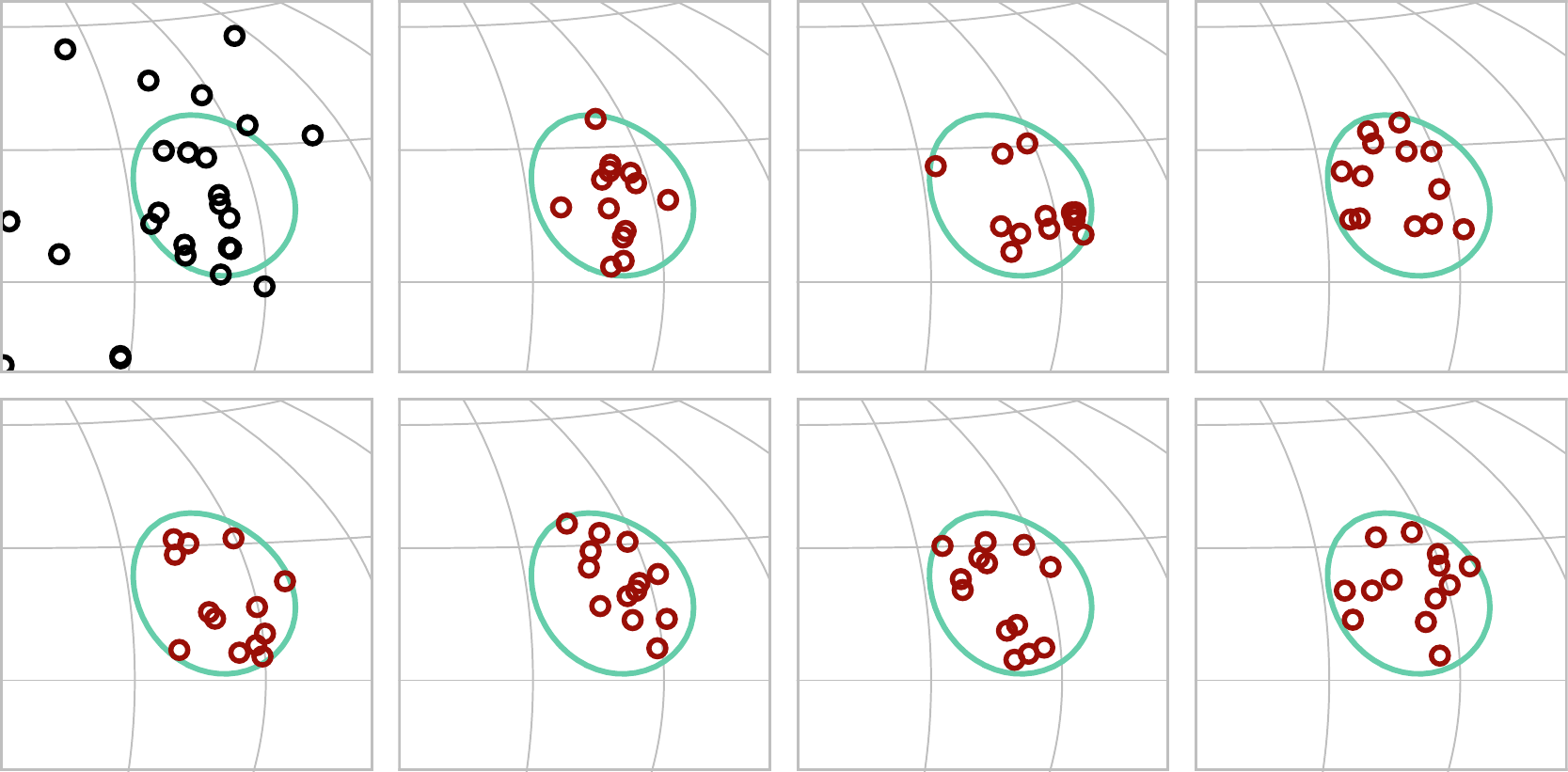}
\caption{Auger data from the region around Cen~A are shown in the top left panel.  This can be compared with simulated sky maps obtained by randomly placing 13 events within a circle of radius $18^\circ$, as shown in the remaining panels.  This results in an average of $\sim 5-6$ pairs within $5^\circ$.
\label{randomness}}
\end{figure}

Since the appearance of multiple images can enable us to directly probe the properties of the local EGMF, it is tempting to attempt to reconcile this with the specific clustering features seen in the Auger data; however, caution is in order given the present statistics. There is a total of 11 pairs of events in the Auger data within $5^\circ$ of each other, 6 of these being within $18^\circ$ of Cen A.  Fig.~\ref{randomness} shows the results from several instances of randomly placing 13 events within a circular region with 18$^\circ$ extent in the sky.  If the signal is distributed roughly uniformly within this region, interesting features can emerge from chance alone.  Considering this, it is difficult to discern between the formation of multiple source images and a scattered signal at this point, although the magnetic field parameters yielding such distributions are not too dissimilar.

An important question is the effect of the Milky Way's magnetic field (GMF) on our results.  It is thought that the GMF consists of a regular component with reversals in the field direction between neighboring arms of the galaxy plus a turbulent component with coherence length of $\sim\,$0.1~kpc~\citep{Stanev:1996qj,Pshirkov:2011um}.  While both have field strengths of a few $\mu$G, the deflection due to the turbulent component is considerably smaller since the regular component is coherent on much larger scales.  Protons with energies of 60~EeV are expected to be deflected by only about a degree (smaller than the uncertainty of UHECR detectors) and less at higher energies~\citep{Tinyakov:2004pw}.

While these deflections could be more important for the regular component of the GMF, two comments are in order: ($i$) The regular component tends to produce only a coherent shift in the position of the source (as can be seen in the sky maps in~\citealt{Stanev:1996qj} and~\citealt{Vorobiov:2009km}) and it cannot account for the fact that the excess events observed are scattered around Cen~A; ($ii$) In at least some of the GMF models, the deflections observed from the direction of Cen~A are less than a degree (in contrast to Galactic Center or disk in which deflections could be much larger)~\citep{Takami:2007kq}.

Since both types of GMF scattering would be significantly larger for heavier nuclei (see, e.g., \citealt{Giacinti:2010dk,Giacinti:2011uj}), it is difficult to reconcile the observed event distribution with an assumption that Cen~A is a source of nuclear primaries.  This, in combination with the fact that Auger does not see an excess of
lower-energy particles from around Cen~A (as discussed above), suggests that if Cen~A is an UHECR source, it is likely producing protons.  We caution here that the larger effects of the GMF on heavy nuclei makes it non-trivial to directly rescale our concluded values for the nearby EGMF.

We thus expect the GMF to have only a small effect on protons at the energies examined here and not to impact our conclusions.  However, it is also possible that some as yet unknown magnetic field component is contributing, which could be even more interesting.  The possible effects of the radio lobes or a field filling an extended halo surrounding Cen~A will be examined elsewhere.

\section{Discussions and Conclusions}
Direct observations have yet to determine the strength of the extragalactic magnetic field near the Milky Way.  While it is difficult to provide details of the field configuration using existing data, we have attempted to determine the most likely values under the assumption that Cen~A is the cause of the UHECR excess coincident with its position in the sky seen by Auger.

While we have only considered a Kolmogorov spectrum here, the overall behavior is essentially determined by the coherence length of the field, so that the effects of using a different power spectrum are small and do not change our basic conclusion that the implied local EGMF is $\gtrsim\,$20~nG.  As seen in Fig.~\ref{limits}, the coherence length is degenerate with the magnitude of the field, thus for smaller coherence lengths even larger fields could be accommodated by the present data.

The presence of a $\gtrsim\,$10~nG field leads to several consequences for all UHECR observations.  In addition to the time delay discussed above, we also see that changes in the particle energy can drastically alter the resulting observed angular distribution.  A general feature that we find is a rapid isotropization with decreasing cosmic ray energy, although specific details vary greatly between both field configurations and the relative positioning of Earth and Cen~A.  Thus inferring the particle spectrum by use of a fixed angular bin can be a non-trivial undertaking.  Unfortunately, we have only a single point of observation, so it is quite possible that our location relative to Cen~A is such that we can observe high-energy events, while not receiving lower-energy cosmic rays, so that a lack of signal is not necessarily unexpected.  This would not be the case if the Galactic fields were solely responsible for the observed dispersion, and it agrees with the lack of UHECR tracks aligned by energy reported by the~\citet{Abreu:2011md}.

Our results confirm that a simple angular projection is not sufficient to arrive at the nature of the local EGMF.  Greater statistics will allow for new diagnostics, such as independently analyzing events from the direction of Cen~A to determine their composition.  If signals of magnetic lensing become evident, this would break the remaining degeneracy in the plane of allowed $B-\Lambda_c$ values in Fig.~\ref{limits}.  One expectation is that events in UHECR clusters caused by lensing should have similar energies, since otherwise a spread would result.  Further, voids of very-low density are visible in Fig.~\ref{augersky}.  If these ``cold spots'' persist, they may yield information about what is {\it not} producing UHECR.

Finally, if Cen~A is not a UHECR source, this would be important to know.  It may be that the excess seen at present, which implies the rather large value of $\gtrsim\,$20~nG, dissipates as more data are obtained.  The direction that future data take will determine what additional steps, such as a greater examination of the effects of magnetic lensing, are warranted.

\acknowledgments
We thank Stirling Colgate, Dan Holz, Hui Li, and especially Matteo Murgia for discussions and comments.
HY is supported LANL LDRD program, TS is supported by DOE Grant DE-FG02-91ER40626, MDK acknowledges support provided by NASA through the Einstein Fellowship Program, grant PF0-110074, and PPK acknowledges support from an NSERC (Canada) Discovery Grant A5713.

\end{document}